\documentclass[ACS,STIX2COL]{WileyNJD-v2}

\articletype{Review Article}%

\received{}
\revised{}
\accepted{}

\begin{document}

\title{The photon-ion merged-beams experiment PIPE at PETRA\,III --- The first five years}

\author[1]{S. Schippers*, T. Buhr,  A. Borovik, Jr., K. Holste, A. Perry-Sassmannshausen}

\author[2]{K.~Mertens, S.~Reinwardt, M.~Martins}

\author[2,3]{S. Klumpp}

\author[2,4]{K. Schubert}

\author[4]{S. Bari}

\author[5,6]{R. Beerwerth, S. Fritzsche}

\author[7]{S. Ricz}

\author[8]{J. Hellhund, A. M\"uller}

\authormark{S. Schippers \textsc{et al}}

\address[1]{\orgdiv{I. Physikalisches Institut}, \orgname{Justus-Liebig-Universit\"at Gie{\ss}en}, \orgaddress{35392 Giessen, \country{Germany}}}

\address[2]{\orgdiv{Institut f\"ur Experimentalphysik}, \orgname{Universit\"at Hamburg}, \orgaddress{22761 Hamburg, \country{Germany}}}

\address[3]{\orgdiv{FS-FLASH-D}, \orgname{DESY}, \orgaddress{22607 Hamburg, \country{Germany}}}

\address[4]{\orgdiv{FS-SCS}, \orgname{DESY}, \orgaddress{22607 Hamburg, \country{Germany}}}

\address[5]{\orgname{Helmholtz-Institut Jena}, \orgaddress{07743 Jena, \country{Germany}}}

\address[6]{\orgdiv{Theoretisch-Physikalisches Institut}, \orgname{Friedrich-Schiller-Universit\"at Jena}, \orgaddress{07743 Jena, \country{Germany}}}

\address[7]{\orgdiv{Institute for Nuclear Research}, \orgname{Hungarian Academy of Sciences}, \orgaddress{4001 Debrecen, \country{Hungary}}}

\address[8]{\orgdiv{Institut f\"ur Atom- und Molek\"ulphysik}, \orgname{Justus-Liebig-Universit\"at Gie{\ss}en}, \orgaddress{35392 Giessen, \country{Germany}}}

\corres{*Stefan Schippers, Justus-Liebig-Universit\"at Gie{\ss}en, I. Physikalisches Institut, Heinrich-Buff-Ring 16, 35392 Giessen, Germany, \email{schippers@jlug.de}}


\abstract[Summary]{The \underline{P}hoton-\underline{I}on Spectrometer at \underline{PE}TRA\,III --- in short, PIPE --- is a permanently installed user facility at the \lq\lq{}Variable Polarization XUV Beamline\rq\rq\ P04 of the synchrotron light source PETRA\,III operated by DESY in Hamburg, Germany.  The careful design of the PIPE ion-optics in combination with the record-high photon flux at P04 has lead to a breakthrough in experimental studies of photon interactions with ionized small quantum systems. This short review provides an overview over the published scientific results from photon-ion merged-beams experiments at PIPE that were obtained since the start of P04 operations in 2013. The topics covered comprise photoionization of ions of astrophysical relevance, quantitative studies of multi-electron processes upon inner-shell photoexcitation and photoionization of negative and positive atomic ions, precision spectroscopy of photoionization resonances, photoionization and photofragmentation of molecular ions and of endohedral fullerene ions.}



\maketitle


\section{Introduction}\label{sec:intro}

\begin{figure*}
\centering \includegraphics[width=0.8\textwidth]{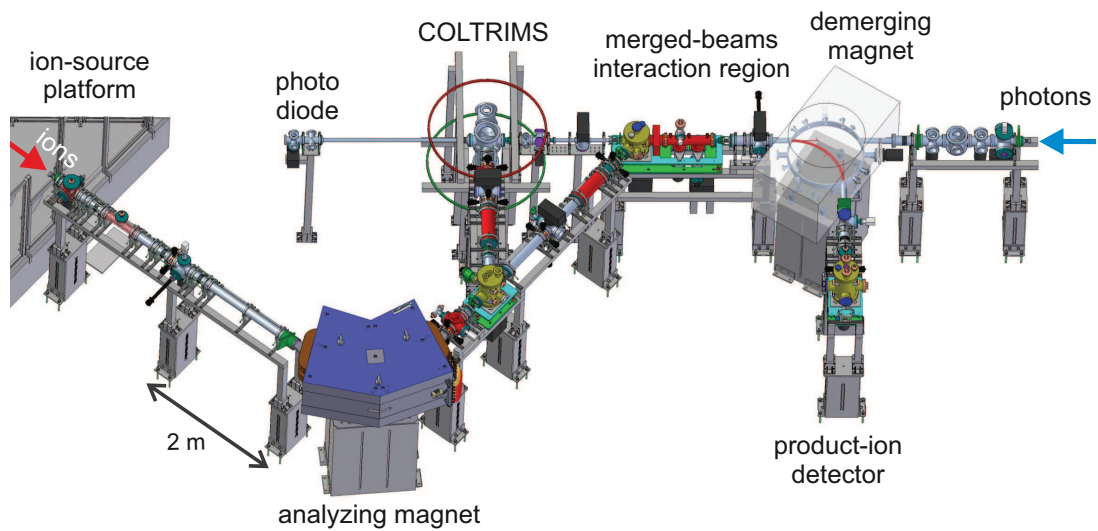}
\caption{\label{fig:PIPE}Sketch of the PIPE setup. The setup has two branches where the ion beam (red arrow on the left) and the photon beam (blue arrow on the right) encounter each other either in crossed-beams or in merged-beams configuration. The crossed-beams interaction volume is surrounded by a COLTRIMS device, results of which (see, e.g., Waitz et al.\citep{Waitz2016a,Waitz2017}) are not discussed here. PIPE has been designed as a user facility. The ions are generated by a user-supplied ion source that is mounted on the ion-source platform. The analyzing magnet serves for selecting the desired primary ion according to its ratio of atomic mass $A$ and ion-charge state $q$. If required, isotopic mass resolution can be achieved. The demerging magnet directs product ions with a given $A/q$ ratio from the photon-ion merged-beams interaction region onto the product-ion detector. Detailed descriptions of all components depicted in this figure and of the experimental procedures can be found elsewhere \citep{Schippers2014,Mueller2017}.}
\end{figure*}

Photoionization experiments with ionic targets \citep{Kjeldsen2006a} are challenging because of the low target densities which are orders of magnitude smaller than the typical densities of neutral gas targets. The photon-ion merged-beams method (see \citep{Schippers2016} for a recent introductory review) compensates the low target density by providing an elongated interaction region of typical $\sim$1~m length where the photon beam and the ion beam move coaxially. In addition,  heavy charged photo products can be detected with nearly 100\% efficiency, since they move with keV energies and can be easily separated from the primary beam by an electric or magnetic field. Nevertheless, the signal rates from such an arrangement are still very small, such that meaningful experiments with synchrotron radiation could only be carried out after the advent of 2$^\text{nd}$-generation synchrotron light-sources. Pioneering work using the photon-ion merged-beams technique was carried out at the Daresbury Synchrotron Radiation Source \citep{Lyon1987a}. Since then, the technique spread to other synchrotron radiation sources, e.g., ASTRID \citep{Kjeldsen1999b}, ALS \citep{Covington2002}, and SOLEIL \citep{Gharaibeh2011a}. Because of their rather large size (heavy magnets and ion sources) these ion-beam setups were realised as permanent installations. Therefore, the available photon-energy range depends on the chosen photon beamline. The latest development is the \underline{P}hoton-\underline{I}on Spectrometer at \underline{PE}TRA\,III (PIPE) \citep{Schippers2014}  which has been set up at the \lq\lq{}Variable Polarization XUV Beamline\rq\rq\ P04 \citep{Viefhaus2013} by a consortium of German university groups.

The PIPE setup (Figure \ref{fig:PIPE}) is unique with respect to the available photon energy range. PIPE is the only photon-ion merged-beams setup where photon energies higher than 1000 eV, i.e., up to currently  2200 eV (up to 3000~eV after completion of the photon beamline) are available. In addition, experiments at PIPE benefit from the record high photon flux that is available at beamline P04. It reaches up to several $10^{14}$~s$^{-1}$ depending on the photon energy $E$ and on the monochromator settings which determine the photon energy spread $\Delta E$. The construction of the PIPE setup was completed in 2012 and the operation of the photon beamline P04 started in 2013. The purpose of this brief review is to summarize the scientific results for positive and negative atomic ions, small molecular ions and endohedral fullerene ions from the first five years of exploitation of the PIPE setup and to provide an outlook on future directions of photoionization experiments with ionic matter in the gas phase.

\section{Atomic ions}

The energy range that is available at PIPE allows one to perform inner-shell ionization studies addressing the $K$ shell of atomic species ranging from carbon to chlorine and the $L$ shell of the elements with nuclear charge up to $Z=44$.  So far, results have been published for $3d$-photoionization of multiply charged Xe$^{q+}$ ions with charge states $q$ in the range $1\leq q \leq 5$ \cite{Schippers2014,Schippers2015a}, for $L$-shell photoionization of Fe$^+$ ions \citep{Schippers2017}, and for $K$-shell photoionization of C$^+$ \citep{Mueller2015a,Mueller2018}, C$^{4+}$ \cite{Mueller2018c}, O$^-$ \cite{Schippers2016a}, F$^{-}$ \cite{Mueller2018b}, and Ne$^+$ \cite{Mueller2017}. The xenon and carbon results have already partly been featured in previous reviews\citep{Schippers2016,Mueller2015}.

\subsection{Responding to astrophysical data needs}

\begin{figure}
\centering
\includegraphics[width=\columnwidth]{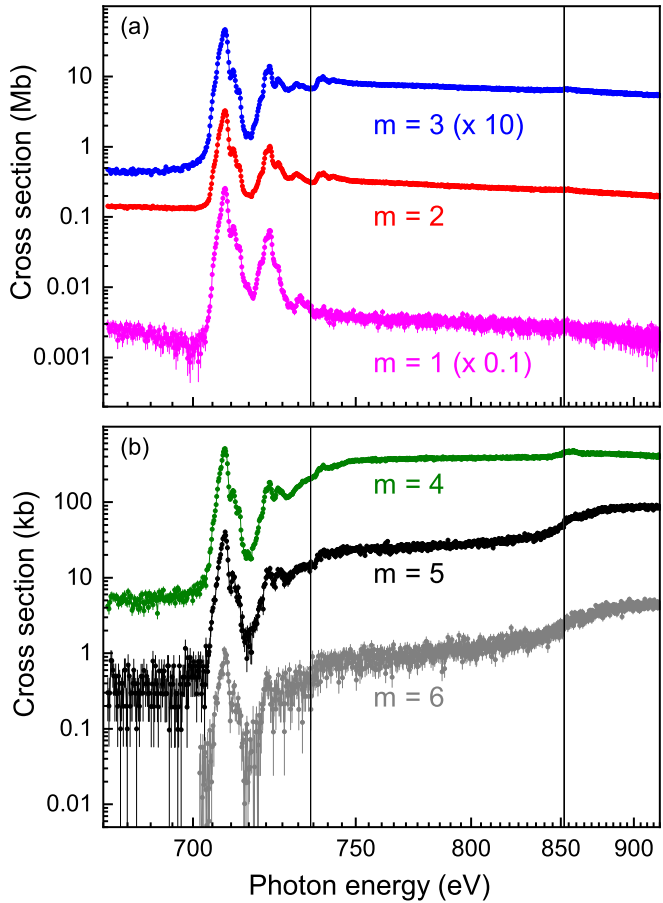}
\caption{\label{fig:Fe}Measured cross sections $\sigma_m$ for $m$-fold photoionization of Fe$^+$ \cite{Schippers2017}. The experimental photon energy spread was about 1~eV and the experimental uncertainty of the photon energy scale amounted to $\pm0.2$~eV. The cross sections $\sigma_1$ and $\sigma_3$ were multiplied by factors of 0.1 and 10, respectively, for a clearer presentation of the data. The vertical lines mark the $2p$ and $2s$ subshell ionization thresholds. It should be noted that the cross-section units are different in panels a) and b).}
\end{figure}

The measurements with carbon and neon ions and particularly the Fe$^+$ experiment were partly motivated by astrophysical data needs. Astronomical x-ray observations of Fe $L$-shell features aim at the detection of iron both in the gas phase and in the solid phase (i.e., dust grains). The spectral features from molecules and solids are expected to differ from those of atoms. Hence, an accurate modeling of the atomic components is critical for inferring the composition of any molecular or solid phase Fe in the interstellar medium (ISM). Near-neutral charge states of Fe are expected to be the dominant gas-phase form of Fe for most regions in the ISM. So far, sufficiently accurate photoabsorption data for an unambiguous identification of these charge states in the ISM have not been available. In response to these data needs, we have carried out measurements of $L$-shell photoabsorption and photoionization data for Fe$^+$, Fe$^{2+}$ and Fe$^{3+}$ ions.

Figure \ref{fig:Fe} shows our published results \citep{Schippers2017} for $m$-fold photoionization ($m=1,…,6$) of Fe$^+$. The experimental photon-energy range 680--920~eV covers the photoionization resonances associated with $2p$ and $2s$ excitations to higher atomic shells as well as the thresholds for $2p$ and $2s$ ionization. The measured cross section values span almost four orders of magnitude ranging from less then 0.1~kb to about 5~Mb. The fact that even the weak cross section for the production of Fe$^{7+}$ could be measured with low statistical uncertainties demonstrates the extraordinary experimental sensitivity of the PIPE setup.

\begin{figure}
\centering
\includegraphics[width=\columnwidth]{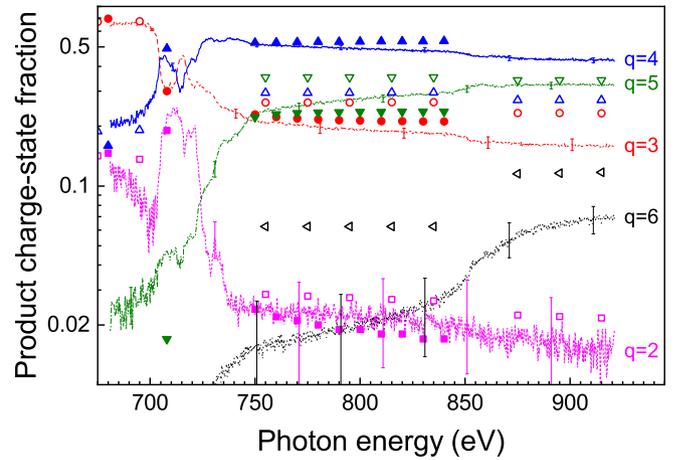}
\caption{\label{fig:Fefrac}Experimental (curves) and theoretical (symbols) product charge-state ($q=m+1$) fractions resulting from $m$-fold photoionization of Fe$^+$ as functions of photon energy \citep{Schippers2017}. The filled symbols represent the fractions obtained from fine-structure resolved cascade calculations \cite{Schippers2017}. The open symbols represent the theoretical results of Kaastra and Mewe \citep{Kaastra1993}.}
\end{figure}

In inner-shell ionization, the initial creation of an inner-shell hole by direct photoionization or by photoexcitation leaves the target ion in a highly excited state which subsequently decays by emitting photons and/or electrons. For many-electron systems, such as low-charge-state iron ions, a large number of deexcitation pathways is available which lead to different final charge states. An accurate determination of the final charge state distribution is important for assessing the charge balance in astrophysical and other plasmas. Theoretical calculations require a detailed handling of complex deexcitation cascades involving radiative and autoionizing transitions. Figure~\ref{fig:Fefrac} presents experimental final charge state fractions resulting from $L$-shell ionization of Fe$^+$ ions as derived from the measured cross sections shown in Figure~\ref{fig:Fe}. The theoretical data were obtained from large-scale cascade calculations \cite{Beerwerth2017,Stock2017} which trace the deexcitation cascades on the fine-structure level. The agreement of the calculated fractions\cite{Schippers2017} with the experimental results is surprisingly good, considering the simplifications that still had to be made to keep the calculations tractable (e.g., shake processes, which have been shown to be important for $K$-shell detachment of the simpler O$^-$ ion \citep{Schippers2016a}, have not been accounted for). The new theoretical results \citep{Schippers2017} are much more appropriate than the results from previous coarser calculations \citep{Kaastra1993} which disregarded fine-structure effects.

\subsection{Quantitative studies of multi-electron processes}

\begin{figure}[bbb]
\centering
\includegraphics[width=0.95\columnwidth]{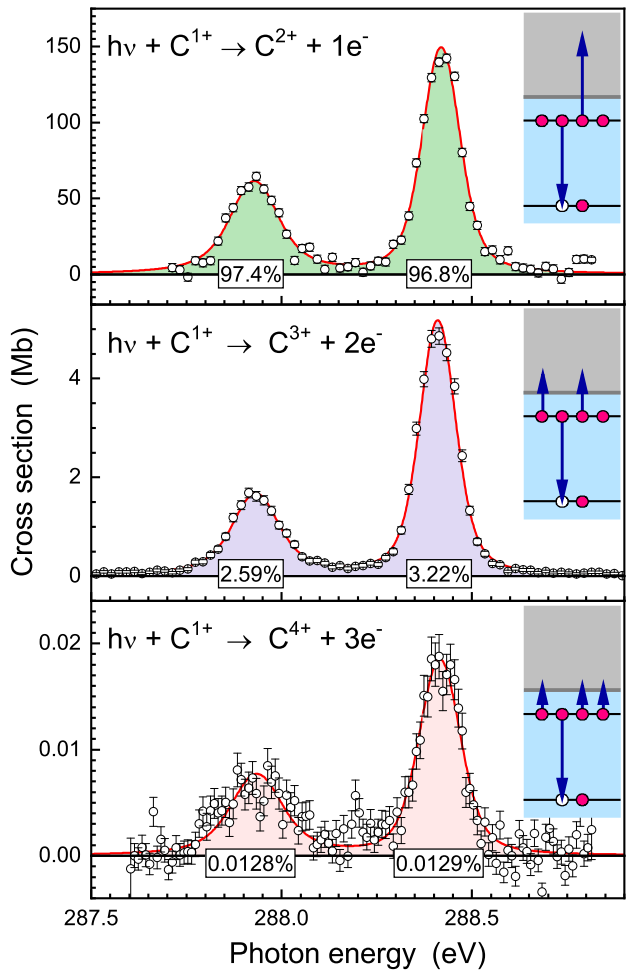}
\caption{\label{fig:multAug}Cross sections for single, double, and triple ionization of ground-state C$^+$ ions\cite{Mueller2015a}. The resonances seen in all the observed channels are associated with $K$-shell excited C$^+$($1s\,2s^2\,2p^2\;^2D/^2P$) terms. The percentages given for the areas of the peaks indicate the branching factors for single, double, and triple Auger decay. The cartoons visualize these processes.}
\end{figure}

\begin{figure}[bbb]
\centering
\includegraphics[width=0.95\columnwidth]{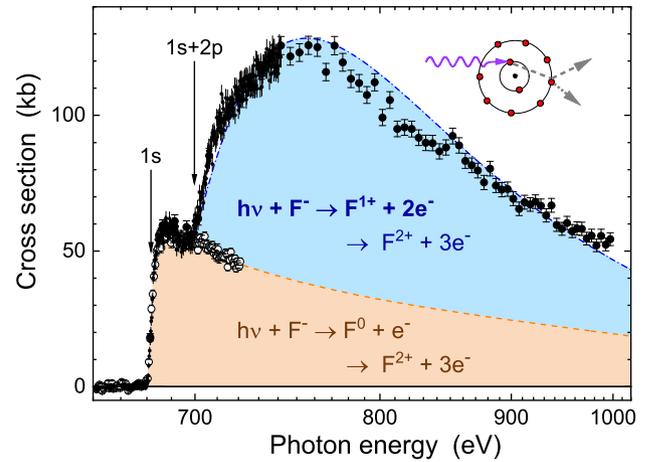}
\caption{\label{fig:Fminus}Measured cross sections for double detachment (open symbols) and  triple detachment (small and large full symbols) of  F$^{-}$($1s^2\,2s^2\,2p^6\;^1S$) ions by single-photon impact \citep{Mueller2018b}. The arrows indicate the thresholds for $1s$-shell ionization and for the simultaneous removal of a $1s$ and a $2p$ electron from F$^{-}$. The triple-detachment cross section is on an absolute scale. The double-detachment cross section has been scaled by a factor 0.153 to match the triple-detachment cross section below $\sim$698~eV and extrapolated (dashed line) to higher energies. The dash-dotted line is the sum of the scaled and extrapolated cross section for double detachment and the scaling\citep{Pattard2002} of cross sections for direct double photoionization. The cartoon visualizes the one-photon--two-electron knock-out process $h\nu + \mathrm{F}^- \to \mathrm{F}^+ + 2e^-$.}
\end{figure}

The already mentioned unparalleled experimental sensitivity of the PIPE setup in combination with the possibility to tailor the electronic structure of the ion under investigation by appropriately selecting its primary charge state allows one to systematically perform quantitative studies of exotic processes which are governed by multi-particle correlations. For example, Müller et al. \citep{Mueller2015a}  prepared  C$^+$($1s\,2s^2\,2p^2$) core-hole levels by $1s\to 2p$ photoexcitation and studied the subsequent decay of these particularly tailored atomic levels by Auger processes. Multiply charged  C$^{q+}$ product ions were detected with charge states $2 \leq q \leq 4$ (Figure \ref{fig:multAug}). Because of the deliberately chosen electronic configuration of the $K$-shell excited C$^+$($1s\,2s^2\,2p^2$) ions with just four $L$-shell electrons, the C$^{4+}$ product ions could only be formed by the simultaneous ejection of three electrons in a triple-Auger process. Since the cross sections for single, double, and triple ionization were measured on an absolute scale, branching ratios for the production of the various final ion charge states (see Figure \ref{fig:multAug}) could be obtained. In addition, the  natural line widths of the photoionization resonances were obtained from high-resolution measurements and, using this additional information, the single, double and triple Auger rates were derived on a purely experimental basis \citep{Mueller2018}. These results have already stimulated new theoretical work \cite{Zhou2016,Liu2018}.

Another multi-electron process that has been addressed at PIPE is the direct knock-out of two electrons by one photon. This fundamental process which is extremely sensitive to
the details of the electron-electron interaction has been a central topic of atomic physics already for decades (see \citep{Mueller2018b} for references). At PIPE, cross sections for double and triple detachment of negatively charged fluorine ions were measured over a considerably larger photon-energy range than in previous studies of direct double-ionization where photon-energies were confined to the near-threshold region.

The results from the PIPE experiment \cite{Mueller2018b} are depicted in Figure \ref{fig:Fminus}. The data exhibit two thresholds at photon energies of 681~eV and 700~eV i) for direct ionization of one $1s$ electron and ii), in case of triple detachment, for direct simultaneous ionization of a $1s$ and a $2p$ electron, respectively. By scaling the double-detachment cross section to the triple-detachment cross section below the second threshold it becomes apparent, that the triple-detachment cross section is essentially made up of two contributions. One component is due to direct ionization of one $1s$ electron and subsequent Auger emission of two electrons. The other component contributes only at energies above the threshold for direct $1s+2p$ double ionization. It is therefore attributed to photodouble detachment (PDD), i.e., direct simultaneous removal of two electrons, followed by a single autoionization event. This interpretation is supported by the fact that the measured cross section complies with the general scaling \cite{Pattard2002} of cross sections for direct double ionization by single-photon impact. The dominance of the PDD contribution to the total triple-detachment cross section is attributed to the small binding energy of the outermost $2p$ electron in the F$^-$ parent ion. The capability of the photon-ion merged-beams method to differentiate between the charge states of the photoions produced subsequently to absorption of a single photon facilitated the clear observation of a process that is characterized by a very small cross section ($< 100$~kb), but dominantly contributes to the production of F$^{2+}$ ions via net triple detachment of F$^-$.

\subsection{Precision spectroscopy of photoionization resonances}

A prerequisite for precision spectroscopy is a low photon-energy spread $\Delta E$ or, in other terms, a high resolving power $E/\Delta E$ of the light source. The monochromator at the photon beamline P04 allows for an resolving power of up to 30\,000 \cite{Viefhaus2013}. This permits measurements of natural line shapes of photoionization resonances. In particular, natural line widths and conversely core-hole lifetimes can be determined reliably.

\begin{figure}
\centering
\includegraphics[width=0.95\columnwidth]{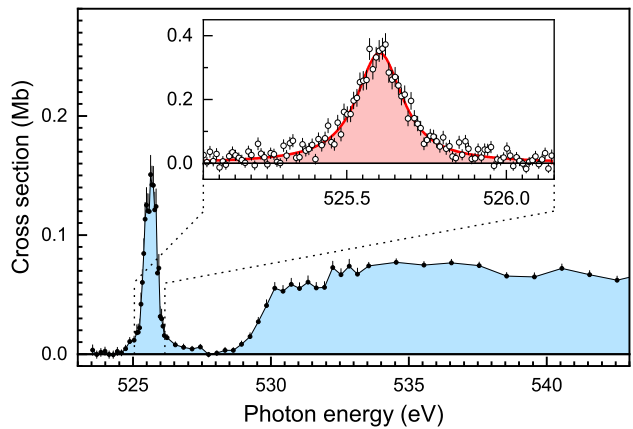}
\caption{\label{fig:Ominus}Measured cross sections for triple detachment of O$^{-}$($1s^2\,2s^2\,2p^5\;^2P$) ions by single-photon impact \citep{Schippers2016a}. The resonance at 525.6~eV is associated with a $1s\to 2p$ excitation to the O$^-$($1s\,2s^2\,2p^6\;^2S$) term. The inset shows a Voigt line profile (full line) that has been fitted to high-resolution experimental data (open symbols, $E/\Delta E \approx 13\,000$).}
\end{figure}

Figure \ref{fig:Ominus} displays the measured cross section \citep{Schippers2016a} for triple detachment of O$^-$ ions in the energy range of the $K$-shell ionization threshold. Negative ions are fundamentally different from neutral atoms or positive ions since the extra electron in a negative ion is not only bound by the long-range Coulomb interaction with the atomic nucleus but, more importantly, also by a short-range attractive force due to the polarization of the atomic core. The accurate theoretical description of these ions still challenges the state-of-the-art quantum computations although the numbers of their bound states are generally finite. A sensitive tool for studying the interactions between the valence and the core electrons is inner-shell ionization of negative ions. Previous studies of $K$-shell photodetachment (e.g.~\citep{Kjeldsen2001a,Walter2006a}) were confined to less complex atomic anions than O$^-$ and considered only double-detachment.

In Figure \ref{fig:Ominus}, the prominent resonance below the $1s$-ionization threshold is associated with a $1s\to2p$ excitation leading to an O$^-$($1s\,2s^2\,2p^6\;^2S$) excited state. From a high-resolution photon-energy scan and a subsequent Voigt line-profile fit to the measured data, the width of this resonance was determined to $164\pm14$~meV corresponding to a $1s$ core-hole lifetime of $4.0\pm0.3$~fs. In addition, systematically enlarged multi-configuration Dirac-Fock calculations \citep{Fritzsche2012a} were performed for the resonant detachment cross sections \citep{Schippers2016a}. Results from these \textit{ab initio} computations agree very well with the measurements for the resonance width and branching fractions for double and triple detachment, if \emph{double} shake-up (and -down) of valence electrons and the rearrangement of the electron density are taken into account.

\begin{figure}
\centering
\includegraphics[width=0.95\columnwidth]{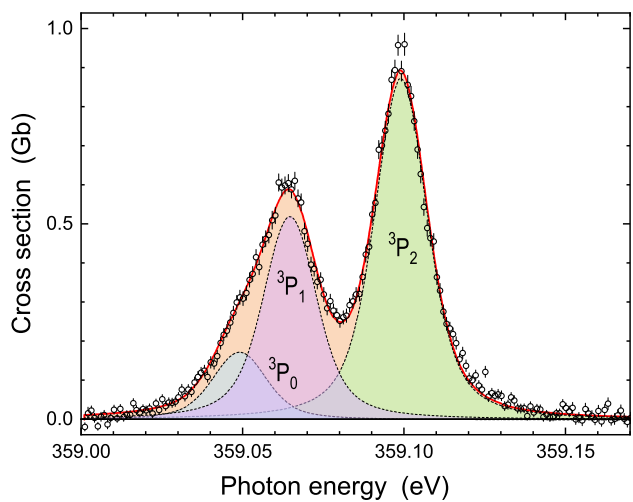}
\caption{\label{fig:C4} Experimental (symbols) and theoretical (lines) cross sections for photoionization of C$^{4+}$($1s\,2s\;^3S$) ions via $2s\,2p\;^3P_{0,1,2}$ resonances \citep{Mueller2018c}. The theoretical cross sections were convolved with a Gaussian with a width corresponding to a resolving power $E/\Delta E = 25\,800$. The individual
contributions of the fine-structure components are shown by thin dotted lines and sum up to the thick full line. The experimental spectrum was shifted up in energy by 1.4~meV to match the theoretical resonance positions.}
\end{figure}

The position of the O$^-$ resonance in Figure \ref{fig:Ominus} was determined to be $525.6\pm0.1$~eV with a statistical uncertainty of only 3~meV. However, a determination of the resonance position with a comparable accuracy was not possible due to the lack of suitable photon-energy calibration standards in the XUV band. The energy scale of the O$^-$ experiment was calibrated against the $1s\to \pi^*$ resonance in molecular oxygen, the position of which is known with a total uncertainty of 90~meV \cite{Prince2003b}.

Photoionization experiments with multiply charged atomic ions have the potential to remedy this unsatisfying situation and to provide much improved photon-energy calibration standards to the soft x-ray community. Figure~\ref{fig:C4} shows a high-resolution scan of the $2s\,2p\;^3P_{0,1,2}$ photoionization resonances produced by $1s\to 2p$ photoexcitation of metastable He-like C$^{4+}$($1s\,2s\;^3S$) ions together with corresponding theoretical results \citep{Mueller2018c}. At a resolving power of 25\,800, the fine structure of this resonance group could be experimentally resolved such that a detailed comparison with large-scale state-of-the-art atomic structure calculations could be made. For atomic systems with only few electrons these calculations, which include, e.g., higher-order quantum electrodynamical (QED) effects, are extremely accurate to at least within 1~meV \cite{Mueller2018c}. Thus, the resulting theoretical resonance energies represent primary reference standards that are two orders of magnitude more accurate than today's standards which are mainly based on electron energy-loss spectroscopy. For the near future, it is planned to transfer these new primary standards to the widely used calibration gases by performing high-resolution scans of  photoionization resonances in He-like and Li-like ions and in atomic and molecular gases with the same settings of the photon beamline.

\section{Diatomic molecular ions}

\begin{figure}[bbb]
\centering \includegraphics[width=0.95\columnwidth]{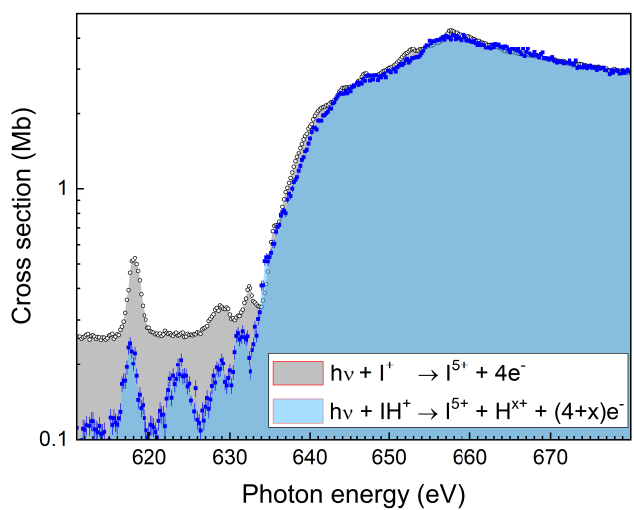}
\caption{\label{fig:IH} Cross sections for the production of I$^{5+}$ ions by photoionization of I$^+$ (open circles) and IH$^{+}$ ions (full squares) in the energy range around the threshold for iodine $3d$ ionization \citep{Klumpp2018}.}
\end{figure}

Molecular ions are of high interest due to the role they generally play in chemistry, e.g., in batteries  or enzymatic reactions.  Moreover, molecular ions have been identified in space where they are created by the impact of cosmic rays or by ultra-violet radiation from nearby stars or other cosmic radiation sources. The number of laboratory studies on the photoionization of ions is limited because the production of ionic targets with sufficient area densities for meaningful photoionization and photofragmentation experiments is challenging. Experimental inner-shell studies with molecular ions are even less in number. In this situation, research has recently focused on simple hydrogen-containing diatomic ions such as the molecular cations CH$^+$, OH$^+$, and SiH$^+$ \cite{Mosnier2016}.

At PIPE, we have extended this sequence to the heavier IH$^+$ molecular ion and performed photon-energy scans at the $3d$ ionization threshold \cite{Klumpp2018}. Figure \ref{fig:IH} shows cross sections for the production of I$^{5+}$ ions by photoionization of atomic I$^+$ ions and by photoionization/photofragmentation of molecular IH$^+$ ions. The atomic cross section is on an absolute scale and the relative molecular cross section was scaled to match the atomic cross section at photon energies above 670~eV. The most obvious differences between the two cross sections concern the resonance structure at energies below 635~eV. The resonances in the atomic spectrum are associated mainly with $3d\to nf$ excitations to higher shells with principal quantum numbers $n\geq 4$, in analogy to what has been discussed in detail for $3d$ photoionization of xenon ions \cite{Schippers2015a}. In the molecular cross section these resonances appear at lower energies. The chemical shifts amount to at most $1.2\pm0.2$~eV in agreement with theoretical predictions \cite{Klumpp2018}. In the molecular spectrum, an additional prominent resonance is observed at 624~eV which has no counterpart in the atomic spectrum. It is associated with the excitation of an iodine $3d$ electron to an unoccupied antibonding $\sigma^*$ molecular orbital in the IH$^+$ ion resulting from hybridization of the iodine $5p_z$ orbital with the hydrogen $1s$ orbital\cite{Klumpp2018}. The resonances appear on a \lq\lq{}background\rq\rq\ cross section due to photoionization of less bound (mainly $4d$) electrons. Apparently, this cross section is higher for I$^+$ as compared to IH$^+$. This difference can be explained by different branching ratios for the production of I$^{5+}$ from I$^+$ on the one hand and from IH$^+$ on the other hand subsequent to electron removal from more loosely bound atomic subshells.

Our IH$^+$ study\cite{Klumpp2018} demonstrates that experiments with molecular ions are feasible at PIPE and that meaningful chemical shifts can be extracted from inner-shell photoionization data.  The accuracy of the  measurements on IH$^+$ ions discussed here was mainly limited by the counting statistics. We are confident that further improvements of our experimental apparatus and, in particular, of our ion-source technology will facilitate even more precise spectroscopic studies of molecular ions at the PIPE setup. In such experiments, the iodine $3d$ lines, e.g., might be used for examining the dynamics of electron excitation in various iodine-containing molecules. Moreover, the merged-beams method holds the promise to yield more state-selective information by monitoring the kinetic energy release upon molecular dissociation \cite{Amitay1996} which we will exploit in future inner-shell x-ray photoabsorption studies with diatomic molecular ions.

\section{Endohedral fullerene ions}

\begin{figure*}
\centering \includegraphics[width=0.75\textwidth]{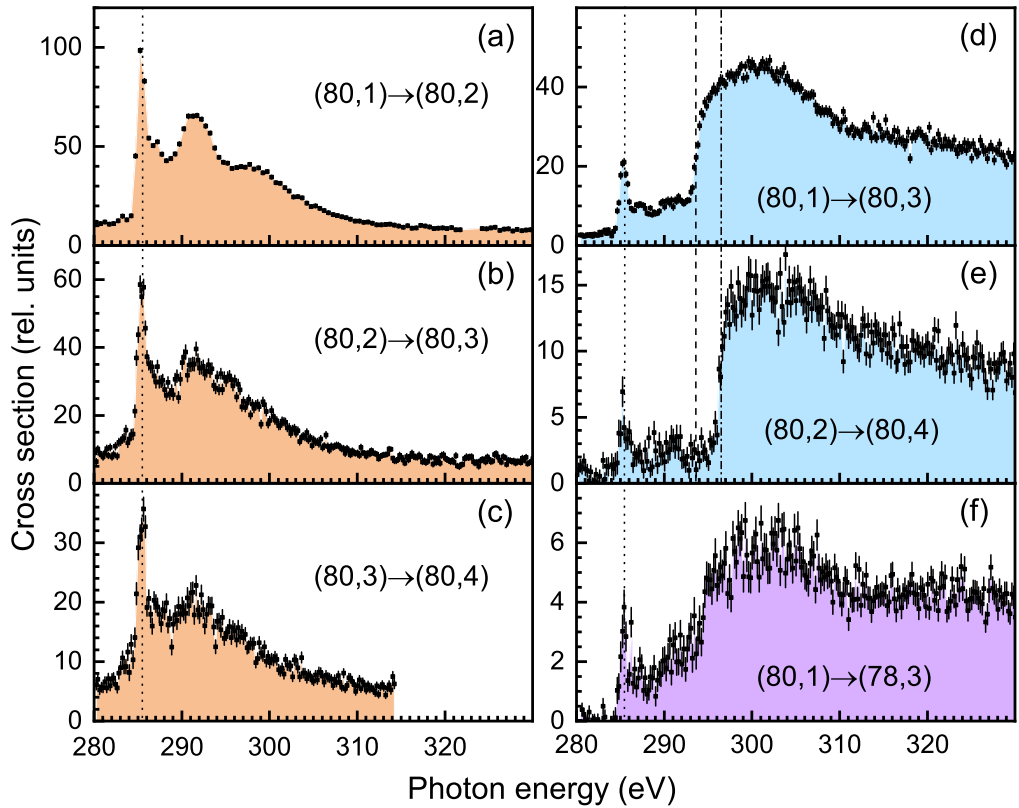}
\caption{\label{fig:endo}  Relative cross sections of photo reactions of singly, doubly, and triply charged Lu$_3$N@C$_{80}$ ions \citep{Hellhund2015,Schippers2016b}.
The short-hand notation ($n$, $q$) $\to$ ($m$, $r$) refers to reactions of the type Lu$_3$N@C$_{n}^{q+}$ $\to$ Lu$_3$N@C$_{m}^{r+}$.
The panels (a) to (c) show single-ionization results and the panels (d) to (f) show double-ionization data. The vertical lines are explained in the text.}
\end{figure*}

Endohedral fullerenes are fascinating objects that have captured the imagination of many scientists\cite{Popov2013}. Their special molecular structure of a closed carbon cage surrounding an encapsulated atom or molecule  has given rise to many intriguing ideas in the basics and applied sciences. Examples are the use of endohedral fullerenes in photovoltaics, quantum computing, medical imaging, or tumor therapy. So far, most of the research on endohedral fullerenes has been theoretical in nature. Experiments were rather limited mainly because the chemical synthesis of endohedral fullerenes is complicated, such that large quantities of high purity material are not readily available. This largely prevented the use of neutral endohedral fullerene vapour as a gas target for, e.g., photoabsorption experiments\citep{Katayanagi2008,Xiong2018}. In contrast to  spectroscopy of a neutral gas target, the photon-ion merged-beams technique permits sensitive measurements of photo-ion yields also when only small amounts of low-purity sample material are available\cite{Phaneuf2013a}.

Previous photon-ion merged-beams experiments with endohedral fullerene ions\cite{Phaneuf2013a,Mueller2007c,Mueller2008b,Kilcoyne2010} addressed outer atomic shells. At PIPE, first results on deep inner-shell photoionization and photofragmentation of Lu$_3$N@C$_{80}$ endohedral fullerene ions were obtained \cite{Hellhund2015}. Figure \ref{fig:endo} shows photo-ion spectra for photon energies in the range 280--330~eV which comprises the threshold for carbon $K$-shell ionization. The ion-yield spectra at the carbon $K$-edge (Figure \ref{fig:endo}) are distinctly different for single ionization on the one hand and for double and triple ionization on the other hand. The single ionization spectra are dominated by resonances that are associated with the excitation of a carbon $K$-shell electron into unoccupied molecular orbitals and subsequent autoionization. As indicated by the dotted line in Figure \ref{fig:endo}, the resonance positions are the same for all product channels of a given primary ion and do not change significantly, either, when going from singly to doubly charged Lu$_3$N@C$_{80}$ targets. In addition, distinct threshold features are observed in the double and triple ionization channels. These are not visible in the single-ionization spectra since the $K$ hole that is created by the ionization event is rapidly filled by a subsequent Auger process leading to the emission of a second electron or even more electrons resulting in double or higher ionization.

A shift of the threshold energy is observed when going from double ionization of the singly to double ionization of the doubly charged ion (dashed and dash-dotted vertical lines in panels (d) and (e) of Figure \ref{fig:endo}). This is due to the additional charge of the product ion and the correspondingly stronger Coulomb attraction of the outgoing photoelectron by the residual ion. This threshold energy shift was used to infer the radius of the fullerene shell and a value of $0.50(4)$~nm was obtained\cite{Hellhund2015}. Within its experimental uncertainty this value is compatible with the range $0.53-0.56$~nm of calculated van-der-Waals radii of $C_{80}$ \cite{Adams1994}.

In addition to the energy range of the carbon $K$-edge, also the energy ranges 390--450~eV and 1500--1700~eV of the N~$K$-edge and the lowest Lu~$M$-edge, respectively, were scrutinized \cite{Hellhund2015}. Ion-yield spectra (which are not displayed here) were taken for heavy photo products ranging from Lu$_3$N@C$_{80}^{3+}$ to  Lu$_3$N@C$_{72}^{5+}$ and Lu$_3$N@C$_{74}^{6+}$. None of the corresponding spectra exhibited any sign of these edges. Most probably, the absorption of an energetic 1600-eV photon by one of the Lu atoms leads to a much more violent fragmentation event such that large fragments cannot be observed. This is interesting, e.g., from a radiobiological point of view and will be more closely investigated in future follow-up experiments where lighter encaged atoms will be used which require less energetic photons for inner-shell absorption.

\section{Conclusions}

In the first five years of its operations, the photon-ion merged-beams setup PIPE at the synchrotron radiation source PETRA\,III has produced a wealth of new experimental data of unprecedented quality on photoionization of atomic, molecular and endohedral fullerene ions. This research has so far resulted in eleven original publications \citep{Schippers2014, Schippers2015a,Hellhund2015,Mueller2015a,Schippers2016a,Schippers2017,Mueller2017,Mueller2018,Mueller2018b,Mueller2018c,Klumpp2018}. These references contain many more details than what could be treated in the present coarse overview.

In summary, brilliant synchrotron radiation has proven to be the key to precision studies of intricate many-electron processes in photon interactions with small quantum systems. Further improvements of the experimental techniques that have been conceived comprise, for example, the further development of ion-source technology, the use of advanced particle detection schemes, the integration of ion trapping techniques, and last but not least an upgrade of the photon source \citep{Schroer2018}.

\section*{Acknowledgments}

The research reviewed here was carried out at the light source PETRA\,III at DESY, a member of the Helmholtz Association (HGF). We would like to thank K.~Bagschik, J.~Buck, L.~Glaser, G.~Hartmann, M.~Hoesch, F.~Scholz, J.~Seltmann, and F.~Trinter for assistance in using beamline P04 as well as  L.~\'Abr\'ok, D.~Bernhardt, A.~Guda, P.~M. Hillenbrand, P.~Indelicato, A.~L.~D.~Kilcoyne, E.~Lindroth,  D.~W.~Savin, R.~A.~Phaneuf, J. Viefhaus, and A. Wilhelm for their contributions to our research at PIPE. Financial support by the German Federal Ministry for Education and Research (BMBF) provided within the \lq\lq{}Verbundforschung\rq\rq\ funding scheme (contracts 05KS7RG1, 05KS7GU2, 05K10RG1, 05K10GUB, 05K13GUA, 05K16RG1, 05K16GUC, and 05K16SJA) and by Deutsche Forschungsgemeinschaft (DFG, projects Mu~1068/22, Schi~378/12, and SFB925/A3) is gratefully acknowledged.
S.B.\ and K.S.\ were supported by the Helmholtz Initiative and Networking Fund through the Young Investigators Program and by DFG (project SFB755/B03).


\end{document}